\documentclass[preprint,aps,showpacs,pre,amsmath]{revtex4}
\usepackage{epsfig}

\begin{document}

\title{Sequential tunneling and shot noise in  \\
ferromagnet/normal-metal/ferromagnet double tunnel junctions}

\author{H. Giang Bach and V. Hung Nguyen \footnote{Corresponding author,
E-mail: hung@iop.vast.ac.vn}}

\address{Theoretical Dept.,  Institute of Physics, VAST P.O.
Box 429 Bo Ho, Hanoi 10000, Vietnam}

\author{T. Anh Pham}

\address{Physics Faculty, Hanoi University of Education
Xuan-Thuy Road, Cau-Giay Distr., Hanoi, Vietnam}

\begin{abstract}
The tunneling through a ferromagnet/normal metal/ferromagnet
double junction in the Coulomb blockade regime is studied,
assuming that the spin relaxation time of electron in the central
metallic island is sufficiently large. Using the master equation,
the current, the tunnel magnetoresistance (TMR), and the current
noise spectrum have been calculated for devices of different
parameters. It was shown that the interplay between spin and
charge correlations strongly depends on the asymmetry of measured
device. The charge correlation makes both the chemical potential
shift, which describes the spin accumulation in the central
island, and the TMR oscillated with the same period as the Coulomb
staircase in current-voltage characteristics. This effect is
smeared by the temperature. The spin correlation may cause an
enhancement of noise at finite frequencies, while the zero
frequency noise is still always sub-poissonian. The gate voltage
causes an oscillation of not only conductance, but also TMR and
noise.
\end{abstract}
\maketitle

\section{Introduction}
Recently, transport in ferromagnetic single electron transistors
(F-SETs) attracts much attention from both fundamental and
application points of view  \cite{review}. In such a device a
small metallic island (or quantum dot (QD) \cite{konig}) is
coupled via tunneling junctions to two leads and capacitively to a
gate. Depending on materials of QD and leads (normal (N) or
ferromagnetic (F) metal) there are three kinds of F-SET: F/F/F-SET
\cite{barnas,karlss,martin}, F/F/N-SET \cite{barnas,fert}, and
F/N/F-SET
\cite{brat,imamu,bulka1,bulka2,korotk,tserk,kuo,braig,wetz}. In
these F-SETs the interplay between spin and charge correlations
during the tunneling process leads to new phenomena such as an
oscillation of the tunnel magnetoresistance (TMR) with increasing
the bias voltage \cite{barnas} or a non-equilibrium spin
accumulation on the normal metallic QD  \cite{imamu}, which in
turn may produce a negative differential conductance (NDC) and an
accompanied shot noise enhancement \cite{bulka1,bulka2}.

The aim of this work is systematically to study the spin-dependent
transport and the shot noise in Coulomb blockade F/N/F-SETs in the
sequential tunneling regime \cite{grabert,averin}. The equivalent
circuit diagram of the device is drawn in the inset of Fig.$1(b)$.
We assume that two leads are made of the same ferromagnetic metal
such as $Ni$, $Fe$ or $La_{0.7}Sr_{0.3}MnO_3 $, and the QD is made
of a normal metal with a sufficiently long spin relaxation time
such as $Al$. The key parameter in the problem is the polarization
defined as $P = (D_M - D_m )/(D_M + D_m )$, where $D_M$ and $D_m$
respectively are the density of states for the majority and
minority spin bands at the Fermi level in leads. Depending on
materials, $P$ is often between $\simeq 0.3 (Ni)$ and $\simeq 0.7
(La_{0.7}Sr_{0.3}MnO_3 $ (see \cite{review} p.149). Such an
asymmetry between two spin bands causes TMR, defined as $TMR =
(R_{AP} - R_P ) / R_P$, where $R_P$ and $R_{AP}$ respectively are
device resistances in the cases, when magnetizations of two leads
are parallel (P) and anti-parallel (AP). Neglecting the Coulomb
charging effects it was shown that \cite{imamu}
\begin{equation}
TMR = P^2 /(1 - P^2 ).
\end{equation}
In order to examine the interplay between spin and charge
correlations, by solving the master equation, we calculate the
current and the TMR for devices different in structure parameters,
including the polarization $P$ and the lead-magnetization
alignment (parallel and anti-parallel), at a large range of bias
voltage and temperature. The obtained results show that the charge
effect not only makes TMR oscillated as the bias increases, but
also considerably enhances it like that observed in F/F/F - and
F/F/N-SETs \cite{barnas}. We also calculate the shot noise power
as a function of the frequency as well as the bias voltage. Though
the shot noise has been very extensively studied in metallic
QD-structures \cite{buttiker,korot,chao,apl,prb}, for Coulomb
blockade F/N/F - single electron tunneling devices, when the
magnetizations in ferromagnetic leads are collinear, we have found
only two works by Bulka et al. \cite{bulka1,bulka2}. However, in
contrast with our F/N/F-SETs of interest, where, as usually, the
normal metallic QD is assumed to be large enough so that there is
an electron thermalization, in these works the QD is assumed to
have regularly separated discrete electron levels \cite{bulka1} or
to have only a single electron level available for the tunneling
process. Such a difference in model should be manifested in the
noise behavior. Our study shows that though the spin correlation
may cause a considerable enhancement of noise at finite
frequencies, including a high peak, the zero-frequency noise is
still always sub-poissonian even in the NDC region. The
super-poissonian noise observed in \cite{bulka1,bulka2} is
essentially related to particular structures of electron levels
modelled. The gate voltage simply leads to an oscillation of the
TMR and the noise with the same period as that for the
conductance.

The paper is organized as follows. Sec.II is devoted to
formulating the problem and presenting fundamental expressions. In
Sec.III we present numerical results of current-voltage (I-V)
characteristics, of TMR and of noise. In this section the effects
of asymmetry, of polarization, of temperature, and of gate are in
detail discussed. Lastly, a brief summary is given in Sec.IV.

\section{ General consideration }
Within the framework of the Orthodox theory \cite{averin} the
state of the F-SET under study is entirely determined by the
number of excess electrons in QD, $n$. For a given $|n>$-state,
the free energy of the system has the form:
\begin{eqnarray}
F(n) &=& (C_2 V/2 - C_1 V/2 + C_g V_g - ne)^2 /2C_t \nonumber \\
&-& (C_1 V^2/4 + C_2 V^2 /4 + C_g V_g^2 )/2 + e( n_1 - n_2 ) V/2,
\end{eqnarray}
where $e$ is the elementary charge, $ C_t = C_1 + C_2 + C_g$, and
$n_1 ( n_2 )$ is the number of electrons that have entered the QD
from the left (right). Any electron transfer across junctions
results in a change in free energy $F$. In the system of interest
there are four possible sequential electron transfers across two
junctions ($\nu$ = 1 and 2), to the right(+) or the left (-). The
change in free energy, $\Delta F_\nu^\pm$, associated with these
electron transfers depends on the relative orientation of
magnetizations in ferromagnetic leads. When the two
lead-magnetizations are in parallel alignment (P-alignment),
$\Delta F_\nu^\pm$ is independent of the orientation of electron
spin [ $\sigma = \uparrow $ (up) or $\downarrow$ (down) ] and can
be directly defined  from (2) as
\begin{equation}
\begin{array}{l}
 \Delta F_1^ \pm  \left( n \right) = e^2 \left( {1 \pm 2n} \right)/2C_t  \mp eC_g V_g /C_t  \mp \left( {C_2  + C_g /2} \right)eV/C_t  \\
 \Delta F_2^ \pm  \left( n \right) = e^2 \left( {1 \mp 2n} \right)/2C_t  \pm eC_g V_g /C_t  \mp \left( {C_1  + C_g /2} \right)eV/C_t  \\
\end{array}
\end{equation}
When lead-magnetizations are in anti-parallel alignment
(AP-alignment), the up (down)-spin electron has a larger (smaller)
tunneling rate for the source junction than for the drain
junction. In this case, to ensure a balance between steady
tunneling currents for each spin orientation, the chemical
potential for up (down)-spin state, $\mu_{\uparrow (\downarrow
)}$, in metallic QD should be shifted upwards (downwards) by (the
same) amount, $\delta \mu_\uparrow  = - \delta \mu_\downarrow
\equiv \delta \mu$. Physically, such a chemical potential shift is
equivalent to a non-equilibrium spin accumulation in QD. Taking
into account this chemical potential shift, the change in free
energy associated with the AP-alignment becomes depending on the
orientation of electron spin $\sigma$ and the chemical potential
shift $\delta \mu$ as :
\begin{equation}
\Delta F_{\nu \sigma }^\pm (n) = \Delta F_\nu^\pm (n) \pm
    (-1)^{\nu - 1} \delta \mu_\sigma \ , \ \nu = 1, 2 ,
\end{equation}
where $\Delta F_\nu^\pm (n)$ is defined in (3) and for the Coulomb
blockade device under study, when the spin-flip relaxation time is
sufficiently large, the chemical potential shift $\delta \mu $ can
be generally calculated from the balanced condition for spin
currents.

Thus, though the free energy $F$ (2) of each charge state is spin
independent, due to a non-equilibrium spin accumulation in QD and
due to a difference between effective tunneling resistances
associated with the majority and the minority spin, the tunneling
rate, in general, depends on the spin orientation of tunneling
electron. For the electron with spin $\sigma$, which tunnels
across the $\nu$-junction, to the right (+) or the left (-), the
rates are given by:
\begin{equation}
\Gamma_{\nu \sigma}^\pm = (e^2 R_{\nu \sigma} )^{-1} \Delta F_{\nu
\sigma}^\pm / [\exp (\Delta F_{\nu \sigma}^\pm / k_B T ) - 1] ; \
\ \ \sigma = \uparrow , \downarrow  \ ,
\end{equation}
where $\Delta F_{\nu \sigma}^\pm$ are corresponding changes in
free energy defined in eqs.(3) and (4), $R_{\nu \sigma } $ depends
on the junction tunneling resistance $R_\nu$ and the spin
polarization $P$ as:
\begin{equation}
R_{\nu \uparrow (\downarrow )} = 2 R_\nu / (1 \pm P_\nu ), \ \nu =
1, 2 ,
\end{equation}
where $P_1 = P_2 \equiv P$ for the P-aligment and $P_1 = - P_2 =
P$ for the AP-alignment device.

Using expressions (3)-(6), in principle, we can solve the master
equation (ME) or perform Monte-Carlo simulations to calculate the
current and further the TMR and the noise. In practice, for simple
structures such as the F-SET under study the ME-method is much
more efficient. Denoting $p(n)$ as the probability of the state
$\mid n>$ of the system, the ME can be written in the matrix form:
\begin{equation}
d\hat{p}(t)/dt = \hat{M} \hat{p}(t) ,
\end{equation}
where $\hat{p}(t)$ is the column matrix of elements $p(n,t)$ and
$\hat{M}$ is the evolution matrix with elements defined as
follows: the diagonal elements,
 $$M(n,n) = -[ \ \sum_{\nu ,\sigma} \Gamma_{\nu \sigma}^+ (n) +
             \sum_{\nu ,\sigma} \Gamma_{\nu \sigma}^- (n) \ ]$$
and non-diagonal ones,
\begin{equation}
M(n,m) = \left \{ \begin{array}{ll}
     \sum_\sigma (\Gamma_{1\sigma}^+ (m) + \Gamma_{2\sigma}^- (m) );
     & \mbox{ \ \ if $m = n - 1$}  \\
     \sum_\sigma (\Gamma_{1\sigma}^- (m) + \Gamma_{2\sigma}^+ (m) );
     & \mbox{ \ \ if $m = n + 1$}  \\
     0; & \mbox{ \ \  otherwise .}
\end{array} \right.
\end{equation}
Solving the ME (7) in the condition $\sum_n p(n,t) = 1$, one can
calculate the stationary currents for each spin channel,
$I_\sigma$, $\sigma = \uparrow$ or $\downarrow$, which are
time-independent and equal to the statistical average of
corresponding currents through any of junctions, $\nu = 1$ or $2$:
$I_\sigma = <I_{\nu \sigma}> = e \sum_n [\Gamma_{\nu \sigma}^+ (n)
- \Gamma_{\nu \sigma}^- (n)] p_{st} (n)$, where stationary
probability $p_{st} (n)$ is defined as $p(n \leftarrow m | \tau
\rightarrow \infty ) = p_{st}(n) \delta_{n m}$ and $p(n \leftarrow
m | \tau )$ is the conditional probability  for having state $|n>$
at the time $\tau$ under the condition that the state was $|m>$ at
an earlier time $t = 0$. This conditional probability obeys the
same ME as for the probability $p(n,t)$. Then, the total
stationary net current is simply given by: $I = I_\uparrow +
I_\downarrow$. Further, from currents associated with the
P-alignment and the AP-alignment of lead-magnetizations, we can
calculate the TMR.

On the other hand, to study the noise we need to know the
time-dependent net current $I(t)$, which for F-SETs has the form:
\begin{equation}
I(t) = \sum_\sigma I_\sigma (t) =
       \sum_\sigma \sum_\nu  g_\nu  I_{\nu \sigma }(t),
\end{equation}
where $I_{\nu \sigma }(t) = e \sum_i [ \Gamma_{\nu \sigma }^+ (i)
- \Gamma_{\nu \sigma }^- (i)] p(i,t)$, \ $g_1 = C_2 / C_t$ and
$g_2 = C_1 / C_t$ \cite{chao}.

In order to calculate the shot noise spectrum $S(\omega )$ of this
current $I(t)$, we extended Korotkov's expression of $S(\omega )$
suggested for metallic SETs \cite{korot} to F/N/F-SETs of interest
and obtain
\begin{eqnarray}
S(\omega ) & = & 2 \sum_\nu g_\nu^2 A_\nu + 4e^2 \sum_{\nu \mu
\sigma \sigma '} g_\nu g_\mu \sum_{n m} [ \Gamma_{\nu \sigma}^+
(n) - \Gamma_{\nu \sigma}^- (n) ]
\nonumber \\
&   & \ \ B_{n m} [ \Gamma_{\mu \sigma '}^+ (m|\mu^- )
p_{st}(m|\mu^- ) - \Gamma_{\mu \sigma '}^- (m|\mu^+ )
p_{st}(m|\mu^+ ) ] .
\end{eqnarray}
Here, $A_\nu = e ( I_\nu^+ + I_\nu^- )$ with $I_\nu^\pm = e
\sum_{n \sigma} p_{st}(n) \Gamma_{\nu \sigma}^\pm $; $\hat{B} =
Re[ (i\omega \hat{I} - \hat{M})^{-1}]$; and $<m|\nu^\pm >$ is the
state obtained from the state $|m>$ by transferring an electron
across the $\nu$-junction to the right (+)/ left (-).

Thus, once the ME has been solved, one can calculate the
stationary current, the TMR and the noise. Except the case of zero
temperature and at low bias of the first Coulomb staircase region,
when the current expression can be analytically derived (see
Appendix), in general, calculations have to be performed
numerically. The results obtained for devices of different
parameters (capacitances, tunneling resistances, polarizations and
magnetization alignments) in a large range of bias voltage and
temperature are presented in the next section, where the gate
effect is also in detail analyzed.

\section{Numerical results}
Numerical calculations have been systematically performed for a
variety of devices with different parameters. The obtained results
will be presented for only typical devices with $(i)$ symmetric
($R_1 = R_2 \equiv 16 R_Q$) and strongly asymmetric tunneling
resistances ($R_1 = 1000 R_Q$ and $R_2 = 5 R_Q$); $(ii)$ two
values of polarization, $P = 0.3$ and $0.9$; and $(iii)$ both P-
and AP-alignments. Here $R_Q = h/e^2$ is the quantum resistance,
and the value $P = 0.9$ is chosen to gain a more clear
manifestation of spin effect. In all devices studied the
capacitances are as follows: $C_1 = 2 C_2 \equiv 2 C$. With chosen
parameters these devices exhibit the most interesting behaviors in
I-V curves such as a clear staircase structure and an NDC.
Regarding the elementary charge $e$, the quantum resistance $R_Q$
and the capacitance $C$ as basic units, in all figures presented
below the voltage, the current, the energy (temperature, chemical
potential) and the frequency are then measured in units of
$e/C$, $e/CR_Q$, $e^2 /C$, and $(CR_Q)^{-1}$, respectively. \\
$(a)$ {\sl Current and chemical potential shift}. Fig.1 presents
I-V characteristics for two typical devices mentioned above,
symmetric $(a)$ and asymmetric $(b)$, in a large range of applied
bias $V$. In each figure two I-V curves corresponding to the case
of P-alignment are certainly coincident (solid lines), independent
of $P$ (0.3 or 0.9). In the case of AP-alignment, on the contrary,
the polarization effect becomes important: the dot-dashed curve of
$P = 0.9$ is much lower than the correspondingly dashed curve of
$P = 0.3$. In addition, for the symmetric device in Fig.$1(a)$ the
difference between two curves, corresponding to the same
AP-alignment, but with different $P$ (or two curves corresponding
to the same $P$ but with different alignments) seems to be much
larger than that for the asymmetric device in Fig.$1(b)$ and
increases with increasing the bias. So, Fig.1 shows a relatively
stronger spin effect in symmetric devices compared to asymmetric
ones. This can be understood from the fact that the charge effect
is more profoundly manifested in the asymmetric device
(Fig.$1(b)$), where I-V curves show clear Coulomb staircases.
\begin{figure}[th]
\centerline{\psfig{file=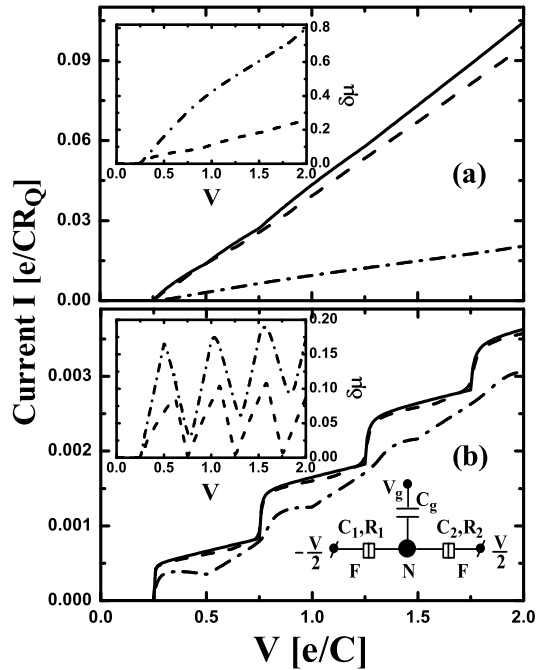,width=7cm}} \vspace*{8pt}
\caption{I-V characteristics for $(a)$ the symmetric device ($R_1
= R_2 = 16 R_Q$) and $(b)$ the asymmetric device ($R_1 = 1000 R_Q$
and $R_2 = 5 R_Q$). For both devices: $C_1 = 2 C_2 \equiv 2 C$.
The currents for P-alignment are independent of $P$ (solid lines).
For the AP-alignment: dashed line - $P = 0.3$; dot-dashed line -
$P = 0.9$. Insets (top and left in both figures): the chemical
potential shift for the same device with the same $P$ as in the
main figure. The equivalent circuit diagram of the device is drawn
in the inset in bottom of Fig.$1(b)$ \ [without gate and $T =
0$].\label{f1}}
\end{figure}

For the AP-alignment, when spin relaxation time is sufficiently
longer than the time between successive tunneling processes, as
mentioned above, a non-equilibrium spin accumulation in the QD may
take place. Such a spin accumulation is equivalent to a difference
between two chemical potentials, corresponding to two spin states.
With increasing the bias, $\mu_\uparrow$ rises to balance the spin
incoming and outgoing rates, and $\mu_\downarrow$ decreases by the
same amount of $\delta \mu $. For systems without Coulomb blockade
this chemical potential shift can be simply evaluated as $\delta
\mu = P e V / 2$ \cite{kuo}, which seems to describe well even the
data in the inset of Fig.$1(a)$ for the symmetric device, when the
linearity of I-V curves are still weakly affected by the charge
correlation. For the asymmetric device, as can be seen in the
inset of Fig.$1(b)$, though the average value of $\delta \mu$ does
increase with increasing $P$ and $V$ similar to that observed in
Fig.$1(a)$, the charge correlation makes $\delta \mu$ oscillated
with the same period of $e/2C$ as the Coulomb staircase in I-V
curves.

In Fig.1 the temperature is zero. Fig.2 describes the finite
temperature effect in the most interesting case of asymmetric
devices with $P = 0.3$ $(a)$ and $0.9$ $(b)$. As the temperature
increases, both the staircase in I-V curves and the oscillation of
chemical potential shifts (see insets) are steadily smeared. In
principle, any finite temperature can destroys the Coulomb gap,
though in practice the current in the gap may be very small when
the temperature is still much smaller than the charging energy.
Remarkably, for the asymmetric device with $P = 0.9$, shown in
Fig.$2(b)$, an NDC region may be appeared at temperatures low
enough. The observed NDC gradually disappears when the temperature
raises. In the particular case of appropriate device parameters at
zero temperature and low bias, when the current expression can be
analytically derived as shown in Appendix, actually, we can also
derive the condition for NDC to be observed.
\begin{figure}[th]
\centerline{\psfig{file=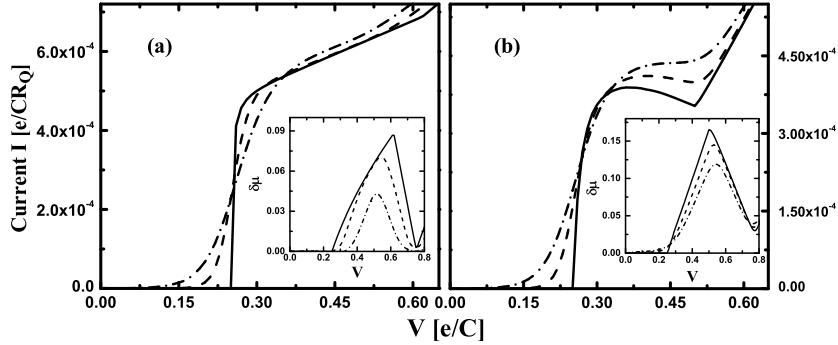,width=11cm}} \vspace*{8pt}
\caption{I-V characteristics for asymmetric devices with $P = 0.3
\ (a)$ and $P = 0.9 \ (b)$ at different temperatures ($k_B T$): 0
(solid line); 0.01 (dashed line); and 0.02 (dot-dashed line).
Insets: the chemical potential shift for the same device and at
the same temperatures as in the main figure \ [without
gate].\label{f2}}
\end{figure}

$(b)$ {\sl Tunnel
magnetoresistance}. As the typical measure of the spin effect in
F-SETs, the TMR attracts much attention from both experimental and
theoretical sides. For F/N/F-SETs under study, as can be seen in
Fig.$3$, both the value and the bias-dependent behavior of TMR are
very sensitive to the asymmetry of SETs [$(a)$ and $(b)$ for
symmetric, while $(c)$ and $(d)$ for asymmetric SETs], to the
polarization [$(a)$ and $(c)$ for $P = 0.3$, while $(b)$ and $(d)$
for $P = 0.9$], and to the temperature [$k_B T = 0$ (solid line),
$ 0.01$ (dashed line) and $0.02$ (dot-dashed lines)]. Overall, in
agreement with those reported for various F-SETs
\cite{barnas,brat,korotk} this figure generally shows that $(i)$
the TMR oscillates with the same period of $e/2C$ as that for
$\delta \mu$ and with an oscillation amplitude decreasing as the
bias $V$ increases; $(ii)$ the TMR generally decreases with
increasing $V$ and becomes constant in the large bias limit; and
$(iii)$ the temperature, smearing the oscillation of TMR at low
biases, does not affect TMR at large biases, when the Coulomb
blockade effect can be neglected. For the symmetric F-SETs, $(a)$
and $(b)$, the large-bias limiting values of TMR, defined quite
well at $eV$ of several charging energy, seem to be in good
agreement with the simple expression (1). For the asymmetric SETs,
$(c)$ and $(d)$, the oscillation of TMR is much stronger and
extends to a larger range of bias. By comparing four these figures
to each other, one can learn much about the interplay between spin
and charge effects in TMR. In particular, the symmetric device
with $P = 0.9$ [Fig.$3(b)$] shows the most remarkable spin effect
with high values of TMR, whereas the charge effect is so weak that
TMR is almost bias-independent except a narrow region close to the
Coulomb blockade threshold voltage.
\begin{figure}[th]
\centerline{\psfig{file=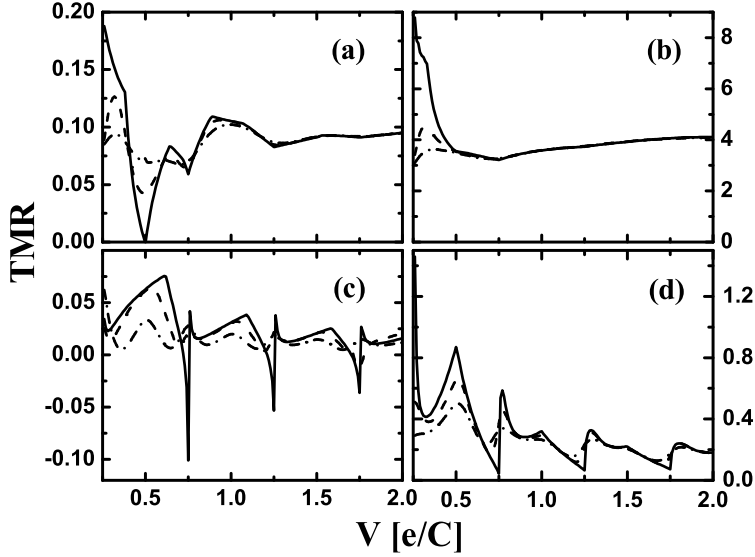,width=10cm}} \vspace*{8pt}
\caption{TMR is plotted as a function of $V$ [$V \ge V_c $, where
$V_c$ is the Coulomb blockade threshold voltage] at various
temperatures ($k_B T$): 0 (solid line); 0.01 (dashed line); and
0.02 (dot-dashed line). Figures: $(a)$ and $(b)$ - symmetric
device with $P = 0.3$ and $0.9$, respectively; $(c)$ and $(d)$ -
asymmetric device with $P = 0.3$ and $0.9$, respectively \
[without gate].\label{f3}}
\end{figure}

Takahashi and Maekawa \cite{takaha} and Kuo and Chen \cite{kuo}
have analyzed the gate effect in the tunneling current and the TMR
in F-SETs. As well-known in the theory of SETs, the gate with
voltage $V_g$ and capacitance $C_g$ leads to the effect similar to
an offset charge of $C_g V_g$, which, in turn, can be related to
the chemical potential shift in QD. Thus, one can expect to see an
oscillation of TMR with increasing $V_g$ like that observed in the
conductance of SETs. In Fig.$4$ we show the TMR versus the gate
parameter $C_g V_g$, calculated for the asymmetric device with $P
= 0.3$ $(a)$ and $0.9$ $(b)$ [the same devices as in Figs.$3(c)$
and $(d)$] at two biases: $V = 0.7$ (solid line) and $V = 1.4$
(dashed lines). It is clear from both figures that all the curves,
though strongly different in value and form, exhibit a
well-defined oscillation with the same period of $e/C_g$ as that
well-known for the conductance. Note that for the device of $P =
0.3 \ (a)$ the TMR though smaller in value is relatively more
sensitive to the change of gate, compared to the device of higher
polarization in figure $(b)$. Moreover, comparing two curves in
each figure, we see that the amplitude of TMR-oscillation is
larger at lower bias, when the Coulomb blockade effect is more
important.
\begin{figure}[th]
\centerline{\psfig{file=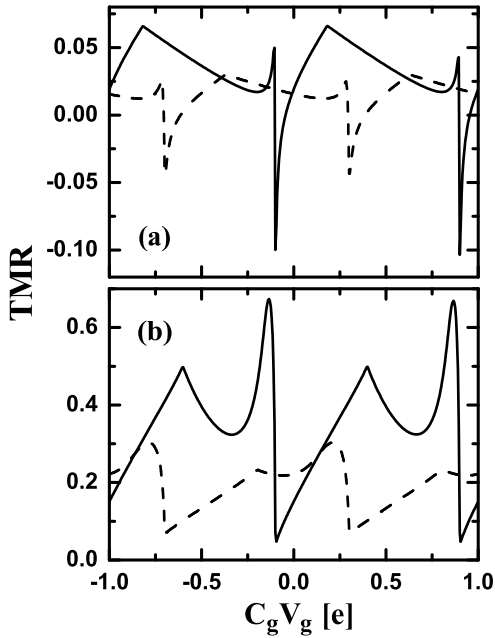,width=6.5cm}} \vspace*{8pt}
\caption{TMR is plotted versus the gate parameter $C_g V_g$ for
asymmetric devices with $P = 0.3 \ (a)$ and $0.9 \ (b)$ at two
biases: $V = 0.7$ (solid line) and 1.4 (dashes line) \ [$T =
0$].\label{f4}}
\end{figure}

$(c)$ {\sl Shot noise}. The shot noise in F-SETs, as
mentioned above, has been studied by Bulka et al.
\cite{bulka1,bulka2}, but for QD-models different from ours. Fig.5
shows the frequency dependence of the normalized noise $S(\omega
)/2eI$ calculated from eq.(10) for the asymmetric devices with
AP-alignment. In each figure [$(a)$ for $P = 0.3$ and $(b)$ for $P
= 0.9$] the results taken at different biases, $V = 0.47, \ 0.53$,
and $0.74$, are compared. These biases are specially chosen from
the typical regions of I-V curve (NDC region, local minimum, and
PDC region). It is clear in both figures that the normalized noise
generally tends to a constant value greater than 0.5 in the limit
of high frequency. This limiting value depends on device
parameters and on bias. It may be even greater than 1 as can be
seen in Fig.$5(b)$. In the opposite region of low frequency, the
picture is more complicated, very sensitive to device parameters.
Even for a given device the noise may experience a very high hill
or a deep valley, depending only on the bias. In fact, from the
noise expression (10) we can at least show there really exists
such a maximum (or minimum) of noise at some frequency. The
calculation, though elementary, is too lengthy to be presented
here. We like only to emphasize that the noise enhancement
observed is due to the spin correlation. This explains why the
noise for the device of $P = 0.9$ in Fig.$5(b)$ is generally
larger than that in Fig.$5(a)$. As one more demonstration, we show
in Fig.$5(b)$ the data for the same asymmetric device of $P =
0.9$, but with P-alignment, and at the same bias $V = 0.74$
(dashed line). It is clear that in this case like in normal
metallic SETs \cite{korot} there is no noise-enhancement in the
whole range of frequency. Here, note that for normal metallic SETs
the high-frequency limiting value of normalized noise is $g_1^2 +
g_2^2$, which is equal to 0.5 for the SET with equal junction
capacitances $[C_1 = C_2]$.
\begin{figure}[th]
\centerline{\psfig{file=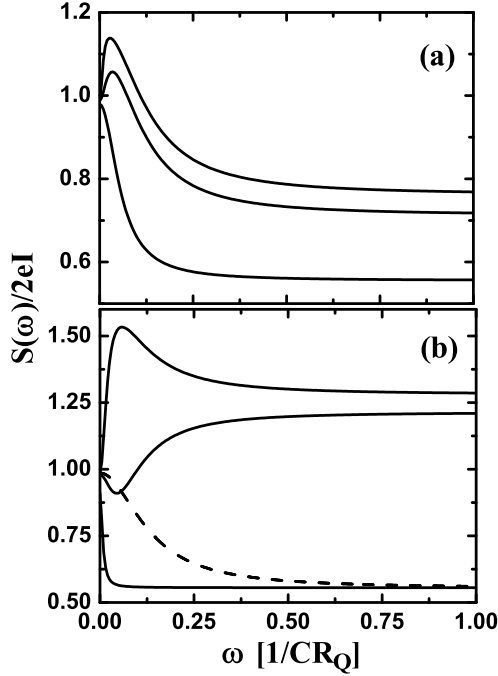,width=6.5cm}} \vspace*{8pt}
\caption{The frequency dependence of normalized noise calculated
for asymmetric devices with AP-alignment and with $P = 0.3 \ (a)$
and $0.9 \ (b)$ at biases (from bottom): $V = 0.49, \ 0.53$, and
0.74. In $(b)$ the data for the same device of $P = 0.9$ at $V =
0.74$, but with P-alignment, is also shown by the dashed line,
which should be compared with the first solid line in the top \
[without gate].\label{f5}}
\end{figure}

In Fig.6 the Fano factor, $F_n = S(0)/2eI$, is plotted as a
function of bias $V$ [two upper curves, see the left axis] for the
same devices as in Fig.5. Interestingly, in both cases under
study, $P = 0.3$ (solid line) and $P = 0.9$ (dashed line), $F_n$
oscillates with the same period of $e/2C$ as that for $\delta \mu
(V)$ [two lower curves, borrowed from Fig.2; see the right axis].
An accurate comparison of two corresponding curves, $F_n$ (upper)
and $\delta \mu$ (lower), for the same $P$ reveals that in each
period the highest value of $F_n$ exactly corresponds to the
minimum in $\delta \mu (V)$. Thus, Fig.6 demonstrates that the
$F_n$- and the $\delta \mu$-oscillations have the same root as the
Coulomb staircase in I-V curves. Experimentally, an exact
correspondence between the noise oscillation and the current
staircase has been observed in the system of a single $InAs$-QD
embedded between resonant tunneling barriers \cite{nauen}. Another
important noise feature deduced from Fig.$6$ is that for
F/N/F-SETs under study the zero frequency noise seems to be always
sub-poissonian [$F_n < 1$] in spite of the noise enhancement
realized at finite frequencies [Fig.$5$] and an NDC region
observed in I-V curves [Fig.2]. Deviations of the shot noise
$S(0)$ from the full (poissonian) value $2eI$, i.e. $F_n \neq 1$,
are the subject of a great number of works [see references in
\cite{prb}]. It is widely accepted that the charge correlation may
suppress or enhance the noise, depending on the conduction regime.
And, two phenomena, NDC and super-poissonian noise [$F_n > 1$] are
not necessarily accompanied by each other \cite{apl,prb}.
\begin{figure}[th]
\centerline{\psfig{file=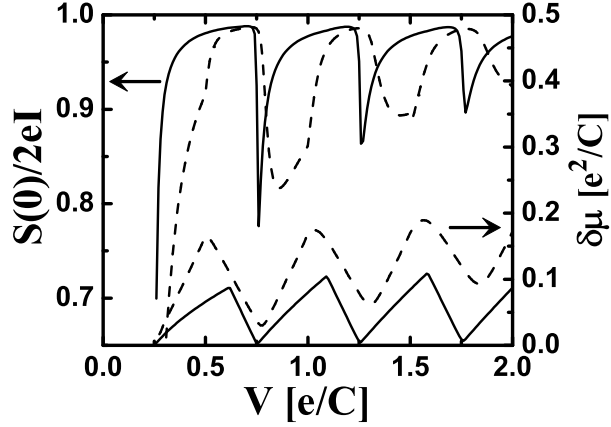,width=8cm}} \vspace*{8pt}
\caption{The zero-frequency normalized noises (two upper curves)
and the corresponding chemical potential shifts (two lower curves)
are plotted versus the bias for asymmetric devices with $P = 0.3$
(solid line)  and $0.9$ (dashed line) \ [without gate].\label{f6}}
\end{figure}

Finally, we demonstrate in Fig.7 how the gate affects the Fano
factor $F_n$ for the same device as in Fig.$6(b)$ [with $P = 0.9$]
and at the same biases [$V = 0.49$ (solid line), 0.53 (dashed
line), and 0.74 (dot-dashed line)] as in Fig.$5$. At any bias, as
expected, the gate makes the factor $F_n$ oscillated with the
period of $e/C_g$ as the gate voltage $V_g$ increases. The
amplitude of oscillation, however, decreases as the bias
increases, showing a gradual weakness of charge correlation
effect. Besides, it is important that, despite the gate-induced
oscillation, the zero-frequency noise $S(0)$ is still always
sub-poissonian.
\begin{figure}[th]
\centerline{\psfig{file=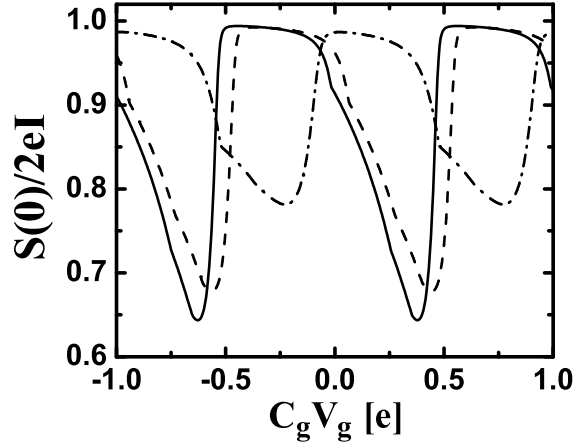,width=7.5cm}} \vspace*{8pt}
\caption{The zero frequency normalized noise is plotted versus the
gate parameter $C_g V_g$ for the asymmetric device with $P = 0.9$
at several biases: $V = 0.49$ (solid line); $0.53$ (dashed line),
and $0.74$ (dot-dashed line).\label{f7}}
\end{figure}

\section{Conclusion}
We have studied the tunneling through F/N/F-SETs in the Coulomb
blockade regime, assuming that the spin relaxation time in the
normal metallic QD is much larger than other characterizing times
in the problem. Using the master equation, we have calculated the
I-V characteristics, the TMR and the current noise spectrum in
devices different in structure parameters, in polarization, and in
lead-magnetization alignment.

It was shown that the interplay between spin and charge
correlations  strongly depends on  the asymmetry of measured
device. While in symmetric devices of equal tunneling resistances
the spin correlation considerably reduces the current, in
asymmetric devices the charge correlation becomes more important,
leading to I-V curves with Coulomb staircases and even with NDC
regions. The charge correlation also makes both the chemical
potential shift, which describes the spin accumulation in QD, and
the TMR  oscillated as the bias increases with the same period as
the Coulomb staircase. All these effects are gradually smeared by
increasing the temperature. In the limit of large bias, when the
charge correlation can be neglected, the TMR becomes independent
of bias. Typical manifestations of the interplay between two
correlations can be also found in the frequency dependence as well
as the bias-dependence of the noise spectrum. While the charge
correlation always suppresses the noise in F-SETs, the spin
correlation may cause a noise enhancement at finite-frequencies,
including a very high peak. The zero-frequency noise is however
always sub-poissonian, regardless of I-V curve behaviors. The gate
voltage causes an oscillation of not only conductance, but also
TMR and noise. While some of current and TMR results obtained in
this work are generally in agreement with those reported in
literature for different kinds of F-SETs, the noise results are
new and useful to better understand the physics of the tunnel
process in
F/N/F-SETs. \\

{\bf Acknowledgments}. We thank V. Lien Nguyen for suggesting the
problem and critically reading the manuscript. This work was
supported by the Ministry of Science and Technology (Vietnam) via
the Fundamental Research Program. \\

\section*{ Appendix }
Following the way suggested in ref.\cite{jpcm} we can derive an
analytical expression for the stationary net current $I$ at zero
temperature within the first Coulomb staircase region, $e/2C_1
\leq V \leq e/C_1$ (assuming $C_1 > C_2$ and $C_g = 0$). Actually,
under these conditions, the rate expression (5) becomes simple as
$\Gamma_{\nu \sigma}^\pm = \Theta (- \Delta F_{\nu \sigma}^\pm
)|\Delta F_{\nu \sigma}^\pm | /e^2 R_{\nu \sigma}$ and all the
probabilities $p(n)$ are equal to zero except those for two states
$|-1>$ and $ |0>$. The master equation can be then exactly solved
that gives
$$I = \frac{[ \Gamma_{1\uparrow}^+ (-1) + \Gamma_{1\downarrow}^+
(-1) ] [ \Gamma_{2\uparrow}^+ (0) + \Gamma_{2\downarrow}^+ (0) ]}
{ \Gamma_{1\uparrow}^+ (-1) + \Gamma_{1\downarrow}^+ (-1) +
\Gamma_{2\uparrow}^+ (0) + \Gamma_{2\downarrow}^+ (0) } . $$
Correspondingly, the quantities $\Delta F_{\nu \sigma}^\pm (n)$ in
the rate expression are now defined as
$$ \Delta F_{1\uparrow (\downarrow )}^+ (-1) = - e^2/2C_t - eC_2
V/C_t \pm \mu  \ ; \ \ \Delta F_{2\uparrow (\downarrow )}^+ (0) =
e^2/2C_t - eC_1 V/C_t \mp \mu \ , $$ where
$$ \mu = [ eV(1 + \gamma ) - \sqrt{ (1 + \gamma )^2 e^2 V^2 - 4(1 -
\gamma )^2 \alpha \beta } \ ] / 2(1 - \gamma ) , $$ $\gamma = (1 -
P)^2 / (1 + P)^2 ; \ \alpha = eC_1 V/ C_t - e^2/2C_t ; \beta =
eC_2
V/C_t + e^2/ 2C_t$ ; and $ C_t = C_1 + C_2$. \\
From the current expression obtained the condition for observing
an NDC could be also derived.
 
\end{document}